\documentclass[12pt]{article}
\usepackage[english]{babel}
\usepackage[dvips]{graphicx}

\title{\bf Optical activity of a neutrino sea in the Standard Model}
\author{G.Karl and V.Novikov\footnote{On leave of absence from the
Institute of Theoretical and Experimental Physics, Moscow, Russia}, \\
Department of Physics, University of Guelph, Ontario, Canada N1G 2W1}
\date{}

\def\fun#1#2{\lower3.6pt\vbox{\baselineskip0pt\lineskip.9pt
\ialign{$\mathsurround=0pt#1\hfil##\hfil$\crcr#2\crcr\sim\crcr}}}

\begin{document}
%\begin{fmffile}{feyn_graph}
\maketitle

\begin{abstract}

We present the Standard Model calculation of the optical
activity of a neutrino sea.

\end{abstract}

\vspace {5mm}

 The idea that intergalactic space is a birefrigent medium for light
due to the presence of a neutrino sea has been contemplated for a
long time. Some thirty years ago, Royer \cite{1} computed in $V-A$
theory an effect of order $G_F$. Later, Stodolsky noted that due to
a theorem of Gell-Mann \cite{2}, there can be no such effect with
massless neutrinos and a point-like coupling, and his observation was
recorded by one of the present authors in a review \cite{3}. In the
early 1980's data on propagation of radio waves through intergalactic
Space put a stringent upper bound on possible optical activity of
the neutrino sea \cite{4} and \cite{5} and this led to renewed
estimates for the size of such effect on the assumption of a neutrino
magnetic moment, which occur for a massive neutrino \cite{5}. More
recently \cite{6} an evaluation was made for a off-mass shell photons
within the Standard Model (SM). Here we note that there is a
frequency dependent effect even for real photons and we also comment
on the effect for virtual photons (both within the SM). Thus we shall
consider three possible effects: a magnetic moment effect, an effect
for virtual photons, an effect for real photons.\\

If the neutrino has a magnetic moment, this moment will interact with
the magnetic field of an electromagnetic wave:

\begin{equation}
\L_{int} = (e{\mu}/2m) \bar{\nu}(p_2) \sigma_{\alpha\beta} q_{\beta}
\nu(p_1) \epsilon(q)
\;\; , \label{1}
\end{equation}

In this equation $\epsilon(q)$ is the polarization vector of a photon
of momenta $q$, while $\nu(p)$ is the neutrino bispinor, and $\mu$
is the neutrino magnetic moment in Bohr magnetons $ e/2m$.\\

With this interaction (1) the forward scattering amplitude of a
photon of momentum $k$ from a neutrino of momentum $p$ is equal to

\begin {equation}
T = 4i(e{\mu}/m)^2  \epsilon_{\mu\nu\alpha\beta} \epsilon_{\mu}(k)
\epsilon_{\nu}^*(k) p_{\alpha} k_{\beta}  \;\;.
\label{2}
\end{equation}

This amplitude differs for left-handed and right-handed photons:

\begin {equation}
 T_{LL} = - T_{RR} = 4(e{\mu}/m_e)^2 (pk)\;\;,
\label{3}
\end{equation}
 and gives rise to a nonzero optical activity for a polarizes neutrino
gas.\\

In the Standard Model the neutrino has no intrinsic magnetic moment,
but a magnetic moment can emerge as a one loop effect if the neutrino
is massive(See figure 1).\\
The order of magnitude of the magnetic moment is:
\begin {equation}
\mu = C (e^2/8{\pi}^2 s^2)(m_e m_{\nu}/m_W^2) \;\; ,
\label{4}
\end{equation}
where $s=sin\theta_W$ and $\theta_W$ is the weak mixing angle;
$s^2 \simeq 0.2312$.
Constant $C$ can be obtained by direct calculation \cite{7}

\begin{equation}
 C =3/4 \;\;.
\label{5}
\end{equation}

Thus  a nonzero optical activity appears in this case as a two-loop
effect
\begin{equation}
 T_{LL}-T_{RR} = (9 e^6/128 \pi^4 s^4) (pk/m_W^2) (m_{\nu}/m_W)^2 \;\;.
\label{6}
\end{equation}

It was understood long ago that ${\gamma\nu}$ scattering takes place
as a one-loop process even for massless neutrino. The first estimates
were performed in the Fermi theory but erroneously \cite{8}.
According to Gell-Mann's theorem \cite{2} the amplitude should
vanish for point-like interactions. Somewhat later Levine \cite{9}
evaluated the scattering mediated by W boson exchange and found a
nonzero amplitude ( See figure 2).
 Due to gauge invariance the amplitude starts with the second power
of the momentum of the photon $k$. Levine computed the first
non-vanishing term, which in the SM looks like:

\begin{equation}
 T = C (e^2/8{\pi}^2 s^2)(pk/m_W^2)^2 \epsilon \epsilon^*\;\;.
\label{7}
\end{equation}

This amplitude is even under parity, in other words it is the same for
right-handed photons and left-handed photons, so it cannot contribute to
optical activity.\\

Very recently it has  been found that by expanding to the next term of
order $k^3$ one finds non-vanishing P-odd terms. The result of this
calculation is \cite{6}:

\begin{equation}
T = (e^4/8{\pi}^2 s^2)\epsilon_{\mu\nu\alpha\beta}
\epsilon_{\mu}(k) \epsilon^*_{\nu}(k) (p_{\alpha} k_{\beta}/m_W^2)
(k^2/3m_e^2)\;\;. \label{8}
\end{equation}

This amplitude vanishes for real photons but is non-vanishing for virtual
photons. The amplitude is singular with respect to the electron mass and comes
entirely from the diagram 2a. \\

It is interesting to ask whether one can get a parity violating amplitude for
real photons. We find a positive answer to this question. In the same order of
expansion over small momenta $k$ ( $k\ll m_W$) one finds the amplitude:

\begin{equation}
T = C(e^4/8{\pi}^2 s^2)(pk/m_W^2)^2\epsilon_{\mu\nu\alpha\beta}
\epsilon_{\mu}^(k) \epsilon_{\nu}^*(k) (p_{\alpha} k_{\beta}/m_W^2)
\;\; ,
\label{9}
\end{equation}
where
\begin{equation}
C  = 4/3 ( \ln(m_W ^2/m^2)-11/3 ) 
\;\;.\label{10}
\end{equation}
This does not vanish for real photons and contributes to the optical activity
of neutrino gas. The logarithmic term in equation (10) comes from diagram 2a,
while the constant receives contributions from all diagrams of figure 2.
The calculation is straightforward.\\

 Before we discuss the order of magnitude
of the different amplitudes we note that the authors of reference \cite{6}
have missed one diagram which contributes to off-shell photons. This is the
diagram with $Z$ exchange shown in Figure 3. For real photons, $k^2=0$ the
triangle diagrams with different fermions inside the loop cancel each other,
so that the Standard Model is anomaly free. But for off-shell photons each
diagram gives a contribution proportional to $k^2/m^2$, where m is the mass
of the fermion running inside the loop  . The main contribution comes from the electron loop and is equal to $(-1/2)$ of the contribution from diagram 2a
 (See ref. \cite{12}) . The sum of contributions from Figure 2a and Figure 3  gives

\begin{equation}
 T = (e^4/8{\pi}^2 s^2)\epsilon_{\mu\nu\alpha\beta}
\epsilon_{\mu}^(k) \epsilon_{\nu}^*(k) (p_{\alpha} k_{\beta}/m_W^2)
(k^2/6m_e^2) \;\;.
\label{11}
\end{equation}

Summarizing these results, we obtain the order of magnitude estimates:
\begin{eqnarray}
 T_{LL}-T_{RR} = \Delta_1 \sim (\alpha^3/s^4)(pk/m_W^2)(m_{\nu}^2/m_W^2)
  for \mu \neq 0, k^2=0 ; \\
\Delta_2 \sim (\alpha^2/s^2)(pk/m_W^2)(k^2/m_e^2)
  for \mu=0, k^2 \neq 0 ; \\ 
\Delta_3 \sim (\alpha^2/s^2)(pk/m_W^2)^3 \ln{m_W^2/m_e^2}
  for \mu=0, k^2=0.\;\; 
\label{12}
\end{eqnarray}

At low frequencies the scattering due to magnetic moment dominates even though it takes
place at two-loop level:
\begin{equation}
 (\Delta_1/\Delta_3) \sim (\alpha/s^2)
(m_W/\omega)^2/\ln{m_W^2/m_e^2} \sim (1.7 GeV/\omega)^2 \;\;.  \label{13}
\end{equation} 
 Here we assume the neutrino at rest, i.e.
$pk=m_ {\nu}\omega $.\\ 

As for the relative value of $\Delta_2$ and  $\Delta_3$

\begin{equation}
 (\Delta_2/\Delta_3) \sim(m_W^2/pk)^2 (k^2/m_e^2) /\ln{m_W^2/m_e^2}
\;\;.
\label{14}
\end{equation}
This is sensitive to the $k^2$ value of the virtual photon. Following the
authors of reference \cite{6} we take for intergalactic space
\begin{equation}
 k^2= \omega_p^2=e n_e/m_e\;\;.
\label{15}
\end{equation}
and assuming $n_e \sim 0.03   cm^{-3} \sim  2.3 \cdot 10^{-43}  GeV^3$, we
obtain:

\begin{equation}
 (\Delta_2/\Delta_3) \sim( 6 keV/\omega)^2 ( 1 eV/m_{\nu})^2\;\;.
\label{16}
\end{equation} 

For visible photons the on-shell activity is smaller than for off-shell
photons, but in the $X$-ray region the reverse is the case. \\
 
  In principle the contribution $\Delta_3$ is similar to the rotation 
from the intergalactic magnetic fields, but the frequency dependence is very
different. A numerical estimate shows that the contribution from the neutrino
sea is many order of magnitude smaller than the magnetic ( Faraday ) effect,
observed with radio waves \cite{10}. We estimate, for radio waves and for $ 1
eV$ neutrino sea with Fermi momenta $k_F\sim 10^{-2} eV$ an effect of order  
$10^{-82}$ $radians/meter$ which is
insignificant compared to the Faraday effect. At shorter wavelengths the
rotatory power due to neutrino sea would increase. These small amplitudes are
matched by very small cross sections for photon-neutrino scattering \cite{11}.
  
  The scattering of laser photons with high energy intensive neutrino 
beam is discussed in the literature ( e.g. see \cite {11} ). For this
case the considered one-loop amplitude dominates over scattering due to 
magnetic moment: 
$ (\Delta_1/\Delta_3) \sim \alpha (m_W^2/pk)^2 (m_{\nu}^2/m_W^2) 
\sim (m_{\nu}/ 1 eV)^2 \cdot(60 keV/\sqrt {s})^4.$

\section{Acknowledgments}

This research was supported by NSERC-Canada and by RFBR
grants 98-02-17372, 98-02-17453 and 00-15-96562.

%\newpage

\newpage
\begin{figure}
\centering
%\begin{tabular}{cc}

%\begin{fmfgraph*}(120,100) 
%\fmfpen{thin}
%\fmfsurroundn{v}{8}
%\fmf{photon,label=$\gamma$,label.side=right}{v3,i5}
%\fmf{fermion,label=$\nu$,label.side=left}{v6,i2}

%\fmf{photon,label=$W$,label.side=left}{i2,i5}
%\fmf{photon,label=$W$,label.side=left}{i5,i1}
%\fmf{fermion,label=$\nu$,label.side=left}{i1,v8}
%\fmffreeze
%\fmf{fermion,label=$e$,label.side=right}{i2,i1}
%\fmfdot{i1,i2,i5}
%\end{fmfgraph*}
%&
%\begin{fmfgraph*}(120,100) 
%\fmfpen{thin}
%\fmfsurroundn{v}{8}
%\fmf{photon,label=$\gamma$,label.side=right}{v3,i5}
%\fmf{fermion,label=$\nu$,label.side=left}{v6,i2}

%\fmf{fermion,label=$e$,label.side=left}{i2,i5}
%\fmf{fermion,label=$e$,label.side=left}{i5,i1}
%\fmf{fermion,label=$\nu$,label.side=left}{i1,v8}
%\fmffreeze
%\fmf{photon,label=$W$,label.side=right}{i2,i1}
%\fmfdot{i1,i2,i5}
%\end{fmfgraph*}
%\\
%\mbox{\rm \bf a)} & \mbox{\rm \bf b)}
%\end{tabular}
\includegraphics{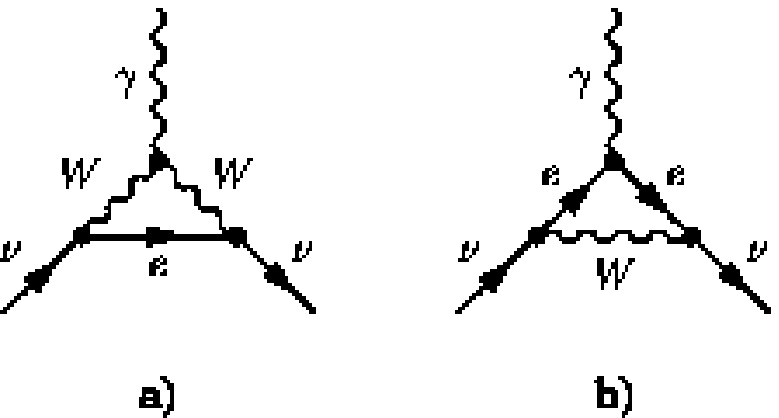}
\caption{Neutrino magnetic moment in the Standard Model.}
\label{fig1}
 \end{figure}

\begin{figure}
\centering
%\begin{tabular}{ccc}
%\begin{fmfgraph*}(120,100) 

%\fmfpen{thin}
%\fmfsurroundn{v}{8}
%\fmf{photon,label=$\gamma$,label.side=left}{v4,i3}
%\fmf{photon,label=$\gamma$,label.side=right}{v2,i4}

%\fmf{fermion,label=$\nu$,label.side=left}{v6,i2}
%\fmf{fermion,label=$e$,label.side=left}{i2,i3}
%\fmf{fermion,label=$e$,label.side=left}{i3,i4}
%\fmf{fermion,label=$e$}{i4,i1}
%\fmf{fermion,label=$\nu$,label.side=left}{i1,v8}

%\fmffreeze
%\fmf{boson,label=$W$,label.side=left}{i1,i2}
%\fmfdot{i1,i2,i3,i4}
%\end{fmfgraph*}
%
%&
%\begin{fmfgraph*}(120,100) 
%\fmfpen{thin}
%\fmfsurroundn{v}{8}
%\fmf{photon,label=$\gamma$,label.side=right}{v2,i4}
%\fmf{photon,label=$\gamma$,label.side=left}{v4,i3}
%\fmf{fermion,label=$\nu$,label.side=left}{v6,i2}

%\fmf{photon,label=$W$,label.side=left}{i2,i3}
%\fmf{photon,label=$W$,label.side=left}{i3,i4}
%\fmf{photon,label=$W$,label.side=left}{i4,i1}
%\fmf{fermion,label=$\nu$,label.side=left}{i1,v8}
%\fmf{fermion,label=$e$,label.side=right}{i2,i1}
%\fmfdot{i1,i2,i3,i4}
%\end{fmfgraph*}
%
%&
%\begin{fmfgraph*}(120,100) 
%\fmfpen{thin}
%\fmfsurroundn{v}{8}
%\fmf{photon,label=$\gamma$,label.side=right}{v8,i1}
%\fmf{photon,label=$\gamma$,label.side=left}{v4,i3}
%\fmf{fermion,label=$\nu$,label.side=left}{v6,i2}

%\fmf{photon,label=$W$,label.side=left}{i1,i2}
%\fmf{photon,label=$W$,label.side=left}{i4,i1}
%\fmf{fermion,label=$e$,label.side=left}{i3,i4}
%\fmf{fermion,label=$\nu$,label.side=left}{i4,v2}
%\fmf{fermion,label=$e$,label.side=left}{i2,i3}
%\fmfdot{i1,i2,i3,i4}
%\end{fmfgraph*}
%\\
%{\rm \bf a)} &  {\rm \bf b)} &  {\rm \bf c)}
%\end{tabular}
%$$
\includegraphics{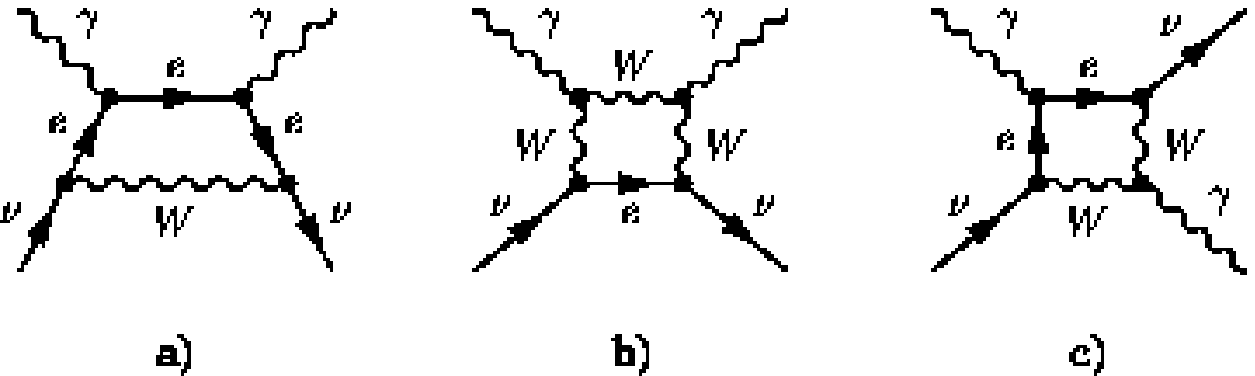}
\caption{Diagrams contributing to $P$-odd $\gamma\nu$ on-shell scattering.}
\label{fig2}
 \end{figure}

\begin{figure}
\centering
%\begin{fmfgraph*}(120,130) 

%\fmfpen{thin}
%\fmfsurroundn{v}{8}
%\fmf{photon,label=$\gamma$,label.side=left}{v4,i3}
%\fmf{photon,label=$\gamma$,label.side=right}{v2,i4}

%\fmf{fermion,label=$\nu$,label.side=left}{v6,i1}
%\fmf{fermion,label=$e$,label.side=left}{i3,i4}
%\fmf{fermion,label=$e$,label.side=left}{i5,i3}
%\fmf{fermion,label=$e$,label.side=left}{i4,i5}
%\fmf{fermion,label=$\nu$,label.side=left}{i1,v8}
%\fmf{photon,label=$Z$,label.side=right}{i1,i5}

%\fmfdot{i1,i3,i4,i5}
%\end{fmfgraph*}
%\\
%$$
\includegraphics{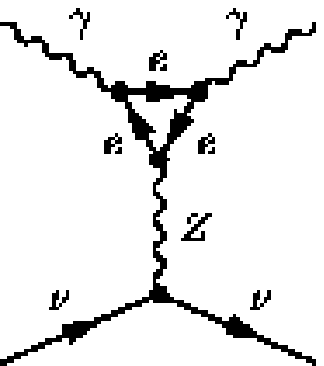}
\caption{Diagram contributing to $P$-odd $\gamma\nu$ off-shell scattering.}
\label{fig3}
\end{figure}
%\end{fmffile}
\end{document}